\documentclass[12pt]{article}
\usepackage{braket}
\usepackage{multirow}
\usepackage{jheppub}
\usepackage{ifpdf}
\usepackage{array}
\usepackage{booktabs}
 
\usepackage{amssymb}
\usepackage{color}
\usepackage{eufrak}
\usepackage{graphicx,psfrag, subfigure}
\usepackage{amsmath}
\usepackage{hyperref}
\usepackage{bbold}
\usepackage{blkarray}
\usepackage[applemac]{inputenc}
\usepackage{dsfont}
\usepackage{yfonts}
\usepackage{setspace}
\usepackage{soul}
\definecolor{dm}{cmyk}{.20, 0, .30, 0}
\sethlcolor{dm}
\usepackage{mathrsfs}
\numberwithin{equation}{section}
 \usepackage{relsize}
\usepackage{tabularx}

\newcommand{\diag}{\text{diag}}

\def\M{M_{{\rm{pl}}}}

\def\be{\begin{equation}}
\def\ee{\end{equation}}
\def\bea{\begin{eqnarray}}
\def\eea{\end{eqnarray}}
\def\calc{\mathsmaller {\mathfrak  C}}

\def\be{\begin{equation}}
\def\ee{\end{equation}}
\def\bea{\begin{eqnarray}}
\def\eea{\end{eqnarray}}

\newcommand{\resc}[1]{{\varpi }_{\cal Q}(#1)}

\begin{document}

\begin{titlepage}

\setcounter{page}{1} \baselineskip=15.5pt \thispagestyle{empty}

\bigskip\
\begin{center}
{\Large \bf Planckian Axions and the Weak Gravity Conjecture}
\vskip 5pt
\vskip 15pt
\end{center}
\vspace{0.5cm}
\begin{center}
{
Thomas C. Bachlechner, Cody Long, and Liam McAllister}
\end{center}\vspace{0.05cm}

\begin{center}
\vskip 4pt
\textsl{Department of Physics, Cornell University, Ithaca, NY 14853 USA}
\end{center}

{\small  \noindent  \\[0.2cm]
\noindent
Several recent works \cite{Tom,Madrid,Madison} have claimed that the Weak Gravity Conjecture (WGC) excludes super-Planckian displacements of axion fields, and hence large-field axion inflation, in the absence of monodromy.
We argue that in theories with $N\gg 1$ axions, super-Planckian axion diameters ${\cal{D}}$ are readily allowed by the WGC.  We clarify the nontrivial relationship between the kinetic matrix  $K$ --- unambiguously defined by its form in a Minkowski-reduced basis --- and the diameter of the axion fundamental domain, emphasizing that in general the diameter is not solely determined by the eigenvalues $f_1^2 \le \ldots \le f_N^2$ of $K$: the orientations of the eigenvectors with respect to the identifications imposed by instantons must be incorporated.  In particular, even if one were to impose the condition $f_N<\M$, this would imply neither ${\cal{D}}<\M$ nor ${\cal{D}}<\sqrt{N}\M$.  We then estimate the actions of instantons that fulfill the WGC.  The leading instanton action is bounded from below by $S \ge {\cal S} \M/f_N$, with ${\cal S}$ a fixed constant, but in the universal limit $S\gtrsim {\cal S} \sqrt{N}\M/f_N$.  Thus, having $f_N>\M$ does not immediately imply the existence of unsuppressed higher harmonic contributions to the potential.  Finally, we argue that in effective axion-gravity theories, the zero-form version of the WGC can be satisfied by gravitational instantons that make negligible contributions to the potential.}

\vspace{0.3cm}

\vspace{0.6cm}

\vfil
\begin{flushleft}
\small \today
\end{flushleft}
\end{titlepage}
\tableofcontents
\newpage
\section{Introduction}\label{intro}
Understanding possible realizations of large-field inflation in string theory is an important problem.  A leading idea is to take the inflaton to be an axion field enjoying perturbative shift symmetries, as in \cite{Natural}.
In axion monodromy scenarios \cite{SW,MSW}, multiple traversals of a sub-Planckian fundamental period lead to a super-Planckian displacement.  In this work we will be concerned with axion inflation scenarios without monodromy, so that large-field inflation is possible only if the fundamental domain for the axion --- or axions --- has a super-Planckian diameter.

Many authors have argued that general properties of quantum gravity should exclude super-Planckian decay constants $f$ for individual axion fields (cf.~\cite{Banks:2003sx}).  Very recently, several papers \cite{Tom,Madrid,Madison} have presented arguments that are rooted in, or parallel to, the zero-form version of the Weak Gravity Conjecture (WGC) \cite{ArkaniHamed:2006dz}, and that claim, with varying levels of finality, to exclude large-field axion inflation in quantum gravity, even in systems with $N>1$ axions.  In this work we will carefully examine the field range limits implied by the WGC \cite{Tom,Madison}, and by the contributions of gravitational instantons \cite{Madrid}.

We will first point out that the diameter ${\cal{D}}$ that is relevant for large-field inflation is {\it{not}} solely determined by the eigenvalues of a matrix of charges, or of the kinetic matrix: as shown in \cite{BLM}, the orientations of the eigenvectors of the kinetic matrix with respect to constraints imposed by periodic identifications strongly affect the diameter.  The problem of computing the diameter amounts to intersecting an ellipsoid of constant invariant distance from the origin with the polytope defining the fundamental domain.  The orientation of the ellipsoid clearly affects the answer.
One consequence is that the limits obtained from the WGC by \cite{Madison}, which take the form of bounds\footnote{The bounded quantities, termed `decay constants' and denoted by $f_n$ in \cite{Madison}, and by $f^{[3]}_n$ here, differ from the quantities we denote by $f_i$ in this work (and in \cite{BLM}): here $f_i$ are the eigenvalues of the kinetic matrix ${\bf K}$, expressed in a Minkowski-reduced basis, cf.~(\ref{Lone}).}
on eigenvalues of a certain matrix of charges, do not imply that ${\cal{D}}<\M$ (nor that ${\cal{D}}<\sqrt{N}\M$).

Next, we consider gravitational instantons in an effective theory of axions coupled to Einstein gravity, as in \cite{Madrid}.  We note that gravitational instantons fulfill the WGC, in the sense that the convex hull of their charge-to-mass vectors $\vec{z}$ contains the unit ball.  We then ask whether such gravitational instantons necessarily contribute unsuppressed higher harmonics to the potential.
In the theory of a single axion with decay constant $f$, one expects gravitational instantons with action $S \sim \M/f$ to make unsuppressed contributions for $f \gtrsim \M$.  Generalizing this expectation to theories of $N$ axions is nontrivial.  The minimum instanton action is determined by the length of the shortest vector in the lattice of charges, with metric given by ${\bf K}^{-1}$.  Surfaces of constant instanton action are ellipsoids determined by the eigenvectors and eigenvalues of ${\bf K}^{-1}$, and in particular the length of the semi-major axis is $f_N$.  In the highly non-generic case where ${\bf K}^{-1}$ is diagonal, so that the ellipsoid is aligned with the lattice, one has $S_{\rm{min}} \sim \M/f_N$.  But a more general possibility is that the ellipsoid does not point directly towards any lattice site, and $S_{\rm{min}}$ can be far larger.  Thus, as in the diameter problem described above, determining the minimum instanton action requires information about the eigenvectors of ${\bf K}$.
We use Minkowski's theorem to obtain an upper bound on $S_{\rm{min}}$, and then argue that this bound is nearly saturated in generic large $N$ theories, with $S_{\rm{min}} \gtrsim \sqrt{N}\M/f_N$.
As a result, in such theories the contributions of gravitational instantons can be neglected even for $f_N \sim \M$.

\begin{figure}
  \centering
  \includegraphics[width=.5\textwidth]{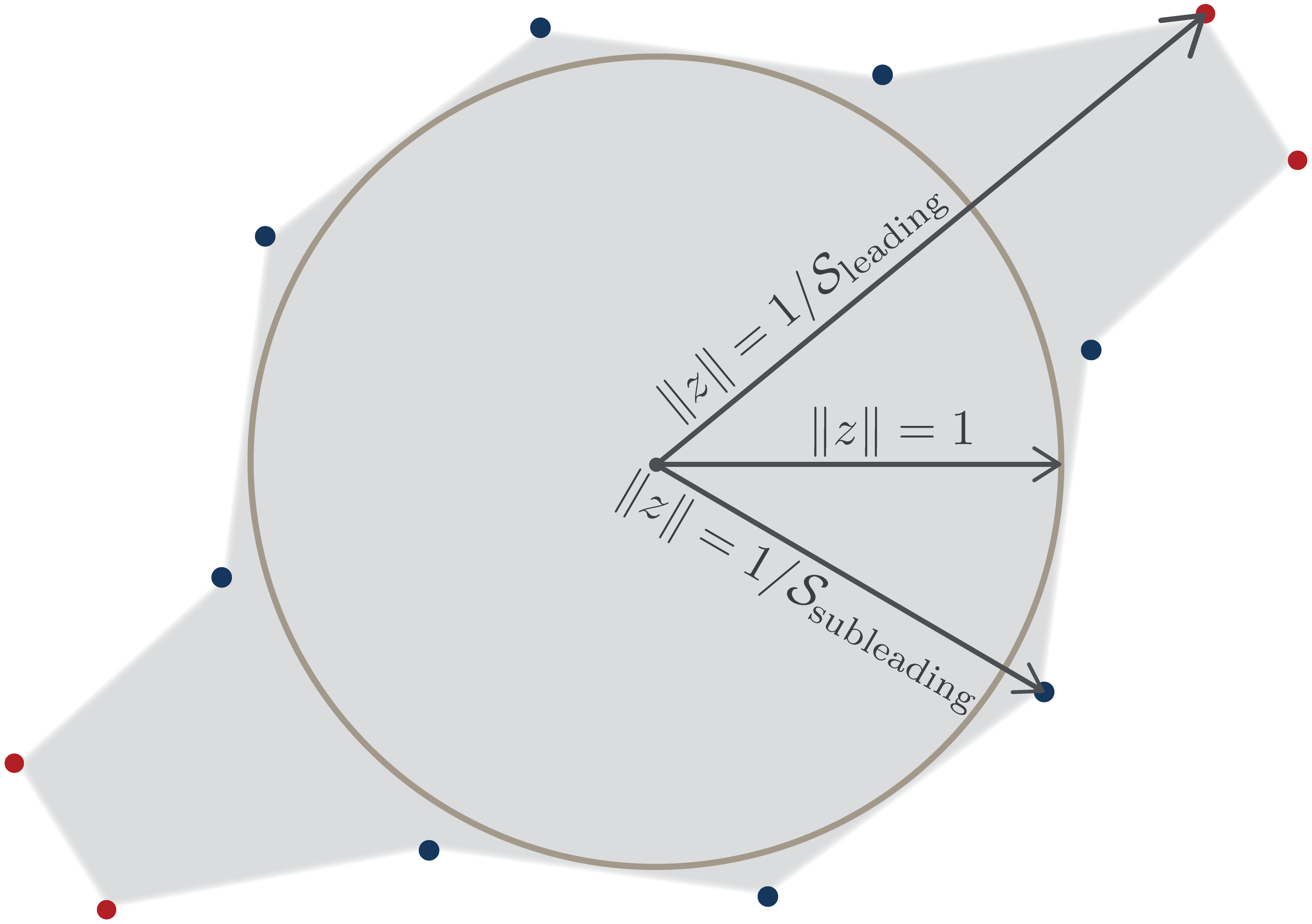}
  \caption{\label{fig:convexhull}\small The unit ball, contained in the convex hull of the charge-to-action vectors of four leading instantons (red) with action prefactor $S_{\text{leading}}$ and eight subleading instantons (blue) with action prefactor $S_{\text{subleading}}$.}
  \end{figure}

The situation is illustrated in Figure \ref{fig:convexhull}: a number of leading instantons give rise to the leading non perturbative potential, but by themselves do not satisfy the convex hull condition. Other instantons do satisfy the convex hull condition but do not contribute significantly to the potential. This scenario was anticipated in \cite{Tom,Madrid,Madison}.

The organization of this paper is as follows.  In \S\ref{sec:2} we describe the geometry of the axion fundamental domain, building on our previous work \cite{BLM}, and carefully explain how to avoid ambiguities that have presented problems in the literature.   In \S\ref{GIWGC} we review the constraints obtained from the WGC and from gravitational instantons, following  \cite{Tom,Madrid,Madison}, and verify that gravitational instantons fulfill the unit-ball form of the WGC.  We combine these ideas in \S\ref{sec:genericactions}, solving the shortest lattice vector problem that determines the dominant instanton, and establishing that higher harmonic contributions to the potential need not be present even when $f_N \sim \M$.  Our conclusions appear in \S\ref{conclusions}.

\section{Geometry of the Axion Fundamental Domain}\label{sec:2}

We will begin by reviewing the geometry of axion field space, following our previous work \cite{BLM}, and precisely defining the diameter of field space that is relevant for assessing whether large-field inflation can occur.  There has been some confusion in the literature because of ambiguities in the definition of an axion decay constant, and because of oversimplified pictures of the relationship between the eigenvalues of the metric on field space, and the diameter of field space.  In this section we will clarify these points in order to lay the groundwork for assessing constraints from the WGC, and from gravitational instantons.  However, the statements in this section are  independent of the WGC, and are deterministic: we defer statements about the {\it{typical}} diameter of field space, as defined in specified ensembles of $N$-axion theories, to \S\ref{sec:genericactions}.

\subsection{Diameter of the fundamental domain}

Consider a theory of $N$ axions $\theta^i$ with discrete shift symmetries $\theta^i\rightarrow \theta^i+2\pi$.
We can write the Lagrangian in a reduced\footnote{A reduced basis consists of $N$ linearly independent, shortest vectors ${\mathscr Q}^i$ in (\ref{canonicalcharges1}) that form a primitive basis: see \S\ref{sec:dominantinstanton}.} basis, including nonperturbative contributions from instantons, as
\be \label{Lone}
{\mathcal L}={1\over 2}K_{ij} \partial\theta^i\partial\theta^j - \sum_{i=1} \M^4 e^{-S^i}\left[1-\cos\left({\cal Q}^i_{\, j}\theta^j \right) \right]\,,
\ee
where $K_{ij}$ is a metric on field space, and ${\cal Q}^i$ corresponds to the {\it{integer}} charge vector of the $i$th instanton with Euclidean action $S^i=S^i_\text{CL}+\delta S^i$, where $S^i_\text{CL}$ is the classical Euclidean action and $\delta S^i$ represents corrections, for example from one-loop determinants.  We order the instanton terms by their size, with $S^1\le S^2\le \ldots$ The charge vectors for canonically normalized fields are then given by
\be\label{canonicalcharges1}
{\mathscr Q}^i={\cal Q}^iS_K \diag(f_i^{-1})\,.
\ee

To understand whether this theory can support large-field inflation, we need to evaluate the maximum invariant displacement that is possible in field space, i.e.~the invariant diameter ${\cal D}$ of the axion field space.  As remarked in the introduction, we will not consider monodromy in this work, so the relevant diameter is that of a {{\it fundamental domain}} for the periodic identifications
\be \Gamma_i: \quad{\cal Q}^i_{\, j}\theta^j \cong {\cal Q}^i_{\, j}\theta^j + 2\pi\,,
\ee
imposed on the axions.  In other words, we are interested in the maximal invariant distance measured along a straight line that does not pass beyond the maxima of any of the cosine terms in (\ref{Lone}).
For the problem of large-field inflation we are only interested in the leading potential contributions; suppose that these are the first $P$ terms of the instanton sum, for some $P\ge N$, and that the remaining terms are small enough to be neglected.
We then ask whether the fundamental domain, now defined by the intersection of the $P$ relevant periodic identifications, admits a super-Planckian diameter.  This question was addressed in \cite{BLM}.

We begin by writing the $P$ relevant entries of $\cal Q$ as
\be  \label{squarechoice}
{\mathcal Q}^i\big|_{i=1,\dots,P}= \left(\begin{tabular}{c} $\mathbf Q$\\ $\mathbf Q_\text{R}$\end{tabular}\right)\, ,
\ee
where $\mathbf Q$ is a full rank set of vectors (see \S\ref{lookout} below for a discussion of ambiguities related to the choice of $\mathbf Q$), and $\mathbf Q_\text{R}$ is a rectangular matrix consisting of all remaining $P-N$ elements. Redefining fields as
\be
\boldsymbol \phi= \mathbf Q \, \boldsymbol \theta\, ,
\ee
we obtain the Lagrangian
\be\label{eqn:lag}
{\mathcal L}={1\over 2}\partial{\boldsymbol \phi}^\top {\mathbf \Xi}\, \partial{\boldsymbol \phi}-\sum_{i=1}^N \M^4 e^{-S^i} \left[1-\cos\left(\phi^i \right) \right]-\sum_{i=1}^{P-N} \M^4 e^{-S^{i+N}}\left[1-\cos\left(\left(\mathbf Q_\text{R} \mathbf Q^{-1} \boldsymbol \phi\right)^i \right) \right]\,,
\ee
where the metric ${\bf \Xi}$ with eigenvalues $\xi^2_N\ge \dots\ge \xi^2_1$ is now given by
\be
\mathbf \Xi= (\mathbf Q^{-1})^\top \mathbf K \, \mathbf Q^{-1}\,.
\ee

The fundamental domain ${\cal M}_{\Gamma}$ is a polytope defined by the $P$ constraints $\Gamma^i$: in terms of the fields $\phi^i$,  ${\cal M}_{\Gamma}$  is an $N$-cube,
\be\label{eqn:constraints1}
-\pi\le  {\boldsymbol \phi}^i \le \pi \, \quad \forall \, i\, ,
\ee
cut by the $2(P-N)$ remaining constraints
\be\label{eqn:constraints2}
-\pi\le \left( \mathbf Q_\text{R} \mathbf Q^{-1}\boldsymbol \phi \right)^i \le \pi \, \quad \forall \, i\, .
\ee
While in general no closed form expression exists for the diameter ${\cal D}$ of ${\cal M}_{\Gamma}$, we obtained a lower bound in \cite{BLM} by computing the diameter along the  direction $\Psi^{\Xi}_N$, corresponding to the eigenvector of ${\bf \Xi}$ with eigenvalue $\xi^2_N$.
Following \cite{BLM}, we can define an operator $\resc{\mathbf w}$ that rescales a vector $\mathbf w$ to saturate the constraint equations~(\ref{eqn:constraints1}) and (\ref{eqn:constraints2}) defining the fundamental domain:
\be\label{rescale}
\resc{\mathbf w}\equiv {2\pi\over \text{Max}_i\left( \{|({\cal  Q} {\bf Q}^{-1} {\mathbf w})_i|\}\right) }\times {\mathbf w}\,.
\ee
We then obtain the diameter \cite{BLM}
\be\label{kinrange}
{\cal D}_{\Psi^{\Xi}_N}=\left \lVert \diag{\xi_i}\, \mathbf S_{\Xi}^\top \resc{\Psi^{\Xi}_N}\right \rVert=\left \lVert \resc{\Psi^{\Xi}_N} \right\rVert  \xi_N\,.
\ee
Evaluating (\ref{kinrange}) in any given example is straightforward, and the form (\ref{kinrange}) will be particularly useful when we study ensembles of theories, with metrics defined by random matrix models, in \S\ref{sec:genericactions}.

\subsection{Eigenvalues vs. the diameter} \label{lookout}

While in this work we will be concerned with the implications of the WGC for general N-axion theories, rather than the construction of individual examples, it may nevertheless be helpful to clarify an apparent conflict in the literature.  Finding an example of an explicit flux compactification in which the axion fundamental domain has a super-Planckian diameter ${\cal D}$ --- which according to \cite{Madison} is incompatible with the WGC --- is not difficult \cite{BLM}. The resolution proposed by \cite{Madison} is that the example of \cite{BLM} must receive corrections from some new source, e.g.~non-BPS instantons omitted in \cite{DDFGK}, that reduce the size of the axion field space in order to make it compatible with the WGC.  But there is a more immediate resolution: the bound obtained in \cite{Madison} from the WGC constrains the largest `axion decay constant' $f^{[3]}_N$, as defined in \cite{Madison}, to obey $f^{[3]}_N< M_p$.  As we explain below, the notion of a decay constant is ambiguous in this context, but let us temporarily grant the proposition that the WGC obliges the quantity $f^{[3]}_N$ to obey $f^{[3]}_N< M_p$.  {\it{This proposition does not imply that the diameter ${\cal D}$ of the axion fundamental domain is sub-Planckian.}}  As such, it does not, in itself, exclude large-field axion inflation, and does not necessitate corrections to \cite{DDFGK} (although quantifying possible corrections would nevertheless be valuable).  In particular, the $f^{[3]}_N$ obtained in \cite{BLM} obeys $f^{[3]}_N = 0.02 \,\M$, easily obeying the bound claimed by \cite{Madison}, while realizing an axion diameter ${\cal D}=1.13\,\M$.

When determining an inflationary trajectory, it is most practical to refer to the invariant diameter of the fundamental axion domain, as computed in \cite{BLM} and reviewed above. This diameter is {\it{not}} given directly by the eigenvalues $f_i^2$ of ${\bf K}$, since this information alone is ignorant of the axion periodicities given by the potential.\footnote{For example, in classes of theories defined in \cite{BLM} we found ${\cal D}\sim N f_N$, which is parametrically larger than the Pythagorean sum of the $f_i$.}  The diameter is also not directly related to the eigenvalues $\xi_i^2$, as the choice of the matrix ${\bf \Xi}$ in (\ref{eqn:lag}) is not unique. The metric ${\bf K}$ is related to the metric ${\bf \Xi}$ by a general linear transformation, so the eigenvalues $\xi_i^2$ can be much larger (in particular, parametrically larger in $N$) than the eigenvalues $f_i^2$. In equation~(\ref{squarechoice}), we made an arbitrary choice of a full-rank matrix $\cal Q$ to set $N$ cosine arguments to be the fields $\phi^i$. The choice of a full-rank square matrix $\mathbf Q$ from the rectangular matrix ${\mathcal Q}$ is not unique in general, {\it{and the metric ${\bf \Xi}$ depends on this choice, while the diameter ${\cal D}$ does not}}.  Thus, even though the first two terms in (\ref{eqn:lag}) could be identified as the Lagrangian of a theory of $N$ axions with precisely $N$ instanton terms, and decay constants $\xi_i$, there are additional terms that contain $P-N$ linear combinations of the fields $\phi^i$. A different choice of ${\bf Q}$ would then yield different $\xi_i$, and therefore the eigenvalues $\xi_i$ are not physical, invariant quantities. Said differently, we have different choices of ${\bf Q}$, some of which may yield very large metric eigenvalues $\xi_i^2$, but only one invariant field space diameter.

The term  ``axion decay constant'' may appear to refer to a well-defined physical quantity, because the relevant notion is unambiguous in the case N=1.   However, in more general settings with $N>1$, and especially with $P>N$, there are significant ambiguities, and it is problematic to refer to metric eigenvalues as ``axion decay constants'', unless these are specified by a unique and invariant definition. Due to the persistent ambiguities in the literature we refrain from using the term altogether.

\section{Gravitational Instantons \& the Weak Gravity Conjecture} \label{GIWGC}

In the previous section we reviewed the geometry of the axion fundamental domain and clarified terminology. We now turn to a precise formulation of the Weak Gravity Conjecture for zero-forms and obtain conditions under which it is satisfied by instantons. We will then estimate the instanton action to verify that instantons that fulfill the WGC need not spoil the flatness of the potential.

\subsection{The Weak Gravity Conjecture}

The Weak Gravity Conjecture \cite{ArkaniHamed:2006dz} is the principle that gravity is the weakest force in a quantum gravity theory.
The weak (or `mild') version of the WGC asserts that there exists a particle  whose charge to mass ratio $Q/M$ exceeds that of an extremal black hole, while the strong version of the WGC asserts that for the lightest charged particle, $Q/M$ exceeds that of an extremal black hole.
The WGC was motivated by a number of ideas about quantum gravity, but most directly emerges from the stipulation that the number of exactly stable particles (in any fixed direction in charge space) should be finite: this requirement implies the mild form of the WGC.  In this work we will assess the implications of the strong and mild forms of the WGC for axion inflation, but provide no new evidence for or against the WGC as a candidate principle.

For our purposes it will be important to extend the logic of \cite{ArkaniHamed:2006dz} to cases with multiple gauge groups, as developed in \cite{Cheung:2014vva}.  Consider a theory containing particle species $i$ with charge vectors $\vec{q}_i$ and masses $m_i$, and define the vector $\vec{z}_i \equiv \vec{q}_i/m_i$, in units where an extremal black hole has $|\vec{z}_{BH}|=1$.  The mild form of the WGC is then equivalent \cite{Cheung:2014vva} to the statement that the convex hull of the vectors $\vec{z}_i$ contains the unit ball.

To generalize the WGC from gauge theories to axions, it is necessary to identify the analogues, in the axion context, of charge and mass.  Roughly speaking, the WGC for a single axion with decay constant $f$ and instanton action $S$ states that $Sf \lesssim \M$ \cite{ArkaniHamed:2006dz}; cf.~\cite{Tom}.
A proposal for a more precise formulation, for axions arising in string theory, was obtained by Brown et al.~\cite{Madison} via T-duality.
In a theory of axions descending from $p$-forms, with charges ${\mathscr{Q}}^i$ and instanton actions $S^i$, mild WGC is the condition that the convex hull of the vectors $z^i \equiv \M{\mathscr{Q}^i}/S_i$ contains the ball of radius $r_{p}$, where $r_p=1$ for $p$ odd, $r_{4}=1$, and $r_{2}=2/\sqrt{3}$~\cite{Madison}.

\subsection{Gravitational instantons}\label{sec:gravinstantons}

Montero et al.~\cite{Madrid} have recently proposed that constraints parallel to, but independent of, the WGC arise from the effects of gravitational instantons.  The claim of \cite{Madrid} is that in theories with sufficiently large super-Planckian axion decay constants, gravitational instanton contributions to the axion potential lead to unsuppressed higher harmonic terms that preclude inflation.  In a single-axion theory this result aligns with earlier arguments \cite{Banks:2003sx} against super-Planckian decay constants.  However, we will show below that in theories of $N \gg 1$ axions, the constraints from gravitational instantons do not present an obstacle to large-field inflation.  Moreover, we will argue that there is a direct connection to the WGC: the convex hull of the charge-to-mass vectors corresponding to gravitational instantons contains the unit ball.  Gravitational instantons therefore ensure that the zero-form WGC is obeyed, and so presents no remaining constraint on non-gravitational instantons, which are generally the dominant terms in the axion potential.

We begin by examining gravitational instantons in low-energy theories of axions coupled to general relativity, closely following \cite{Madrid} and \cite{ArkaniHamed:2007js}, in order to obtain the correctly normalized gravitational instanton action expressed in terms of integer axion charges.  Consider, therefore, the Euclidean Einstein-Hilbert action coupled to a real Euclidean pseudoscalar:
\be \label{EHaction}
S = \int d^4x \sqrt{|g|}\left(-\frac{\M^2}{2}R + \frac{f^2}{2}g^{\mu \nu}\partial_\mu \phi \partial_\nu \phi \right).
\ee

The goal is to find a solution of the Euclidean equations of motion following from (\ref{canonicalcharges}), with the asymptotic profile
\be
\phi(r) \to \frac{n}{4\pi^2 f^2}\frac{1}{r^2}\,.
\ee
To determine the quantization condition on the parameter $n$, we follow \cite{Madrid} and consider expanding the axion action about a charge-$n$ instanton solution $\phi = \braket{\phi_n} + \varphi$:
\be
S_{\phi} = \frac{f^2}{2}\int\left( d \braket{\phi_n}  \wedge \ast d \braket{\phi_n} + d \varphi  \wedge \ast d \varphi + d \braket{\phi_n}  \wedge \ast d \varphi + d \varphi\wedge \ast d \braket{\phi_n} \right).
\ee
The interaction of a charge-$n$ instanton with the axion is
\be
S_{n} = f^2\int d\varphi \wedge \ast d\braket{\phi_n} = f^2\int \varphi  \left(d\ast d\braket{\phi_n}\right)\,,
\ee
where in the last equality we have dropped a total derivative. Now consider shifting $\varphi$ by a constant $\alpha$. The change in the action is
\be
\Delta S_n = f^2\alpha \int d\ast d \braket{\phi_n}\,.
\ee
Applying Stokes' theorem yields
\be
\Delta S_n = f^2\alpha\int\limits_{S^3}\ast d\braket{\phi_n} = \alpha \int \frac{n}{2\pi^2} d\Omega_3 = n\alpha\,.
\ee
We conclude that the potential for the axion has period $2\pi/n$: $V_n \sim \text{cos}\left(n\phi\right)$, and so $n$ can be identified as an integer charge.

With a spherically symmetric ansatz for the metric,
\be
ds^2 = {\mathfrak f}(r)dr^2 + r^2 ds_{S^3}^2\,,
\ee
the Einstein equations have the wormhole solution \cite{Madrid}
\be
{\mathfrak f}(r) = \frac{1}{1-\frac{a}{r^4}}, \quad d\phi(r) = \frac{n}{2\pi^2 f^2}\sqrt{{\mathfrak f}(r)}\frac{dr}{r^3}, \quad a \equiv \frac{n^2}{3\pi^3}\frac{G}{f^2}\,.
\ee
Tracing the Einstein equations we have $ R =- 8\pi G T$, where $T=f^2(\partial \phi)^2$.
We can now evaluate the action. Note that the instanton action will be half of the action quoted above, since a wormhole can be thought of as an instanton---anti-instanton pair. The instanton action is therefore
\be
S^{\text{inst}} = \frac{1}{2}\int d^4x \sqrt{|g|}\left( -\frac{\M^2}{2}R + \frac{f^2}{2}g^{\mu \nu}\partial_\mu \phi \partial_\nu \phi\right) = \frac{f^2}{2} \int d^4x \sqrt{|g|} (\partial_\mu \phi \partial^\mu \phi)\,.
\ee
Integrating over the $S^3$ yields a factor of $2\pi^2$. The radius of the instanton $a^{1/4}$ provides a small-length cutoff of the solution, so the action becomes
\be
S^{\text{inst}} = \pi^2 f^2 \int\limits^{\infty}_{a^{1/4}} {\mathfrak f}(r)^{-1/2} r^3 (\phi'(r))^2 dr = \frac{n^2}{4\pi^2f^2}\int\limits^{\infty}_{a^{1/4}} \frac{\sqrt{{\mathfrak f}(r)}}{r^3} dr = \frac{\sqrt{3\pi}n}{16f\sqrt{G}}\,.
\ee
The reduced Planck mass is given by $1/\M = \sqrt{8\pi G}$, so we can finally write the instanton action as\footnote{This result is half as large as the action given in \cite{Madrid}; the difference stems from what appears to be an error in the second equality in their (3.13).}
\be
S^{\text{inst}} = \frac{\sqrt{6}\pi n}{8}\frac{\M}{f}\,,
\ee where as noted above, $n \in \mathbb{Z}$. In the case of $N > 1$ axions, the action takes the form
\be \label{EHactionN}
S = \int d^4x \sqrt{|g|}\left( -\frac{\M^2}{2}R + \frac{1}{2}g^{\mu \nu}\partial_\mu \vec{\phi} K \partial_\nu \vec{\phi}\right)\,,
\ee
where ${\bf K}$ is the metric on the field space. The solution of the Einstein equations takes the same general form, with the modifications
\be
a = \frac{\vec{n}K^{-1}\vec{n}^\top}{3\pi^2}G, \quad \partial_r \vec{\phi} K \partial_r \vec{\phi} = \frac{\vec{n} K^{-1}\vec{n}^\top}{4\pi^4}\frac{{\mathfrak f}(r)}{r^6}\,.
\ee
Evaluation of the  gravitational instanton action with integer charges $\vec{n}$ yields
\be \label{ginstN}
S^{\text{inst}} = \frac{\sqrt{6}\pi }{8}\M\sqrt{\vec{n}K^{-1}\vec{n}^\top}\,.
\ee

\subsection{The convex hull condition for gravitational instantons}

Let us now consider the convex hull of the instantons with classical action of the form
\be
S^i_\text{CL}={\mathcal S}\M\sqrt{({\cal Q}^i) {\bf K}^{-1} ({\cal Q}^i)^\top}\,,
\ee
where ${\mathcal S}=\sqrt{6}\pi/8\approx 0.96$ for gravitational instantons. The zero-form version of the weak gravity conjecture now requires that the convex hull of the vectors
\be
z^i={{\mathscr Q}^i\over S^i}\M
\ee
contains the unit ball. With the charges (\ref{canonicalcharges1}) we have
\be
z^i={{\cal Q}^iS_K \diag(f_i^{-1})\over {\cal S} \sqrt{({\cal Q}^i) {\bf K}^{-1} ({\cal Q}^i)^\top}}\,,
\ee
where ${\cal Q}^i$ are integer vectors. Evaluating the norm $\left \lVert z^i\right \lVert$ we immediately find
\be
 \lVert z^i \rVert={1\over {\cal S}}\,.
\ee
The norm of the vector $\left \lVert z^i\right \lVert$ is independent of the charges, so if the charges are unconstrained integers, the convex hull of the charge to action vectors always encloses a sphere of radius $1/{\cal S}$. The charges of the gravitational instantons considered in \S\ref{sec:gravinstantons} are arbitrary integers and therefore satisfy the convex hull condition of the zero-form version of the WGC.

From the preceding arguments we cannot conclude that in string theory, gravitational instantons fulfill the WGC.  First of all, it is not clear that nonsingular wormholes analogous to those described in \S\ref{sec:gravinstantons} arise in the low-energy effective theories arising from string theory, which generally include additional fields, such as the dilaton --- see \cite{ArkaniHamed:2007js}.  If corresponding gravitational instantons do exist, their actions may have ${\cal{S}} \neq \sqrt{6}\pi/8$, and in particular could have ${\cal{S}}>1$.\footnote{Even for ${\cal{S}} = \sqrt{6}\pi/8$, gravitational instantons cannot fulfill the convex hull condition with radius $r_2=2/\sqrt{3}$ \cite{Madison} that pertains to two-forms in four dimensions, because $\sqrt{6}\pi/8<3/\sqrt{2}$.}

\section{Instanton actions in generic large $N$ theories }\label{sec:genericactions}

In the preceding sections we have emphasized that in theories with more instanton contributions than axions ($P>N$ in our notation), one cannot determine
the diameter of the axion fundamental domain solely from the eigenvalues of the kinetic matrix; one also requires information about the orientation of the eigenvectors of the kinetic matrix with respect to the boundaries resulting from periodic identifications.  When the number of significant instanton contributions to the potential is large, computing the diameter becomes difficult.

At the same time, it is precisely in the case $P>N$ that the WGC fails to constrain  axion alignment mechanisms \cite{KNP,Nflation,Bachlechner:2014hsa} that could give rise to super-Planckian diameters in theories obeying the WGC \cite{Tom,Madison}.  The idea, which we will make precise below, is that the convex hull condition resulting from the WGC may be fulfilled by instantons beyond the first $N$, therefore rendering the first $N$ instantons terms unconstrained by the WGC.
In such a case one could say that the WGC is neutralized by (some of) the $P-N$ additional instanton terms, as anticipated in \cite{Tom,Madrid,Madison}.

However, the additional instantons that fulfill the WGC might introduce leading order terms in the potential, producing new hyperplane-pair boundaries of the fundamental domain, and reducing the diameter.   In this setting, one might expect ---  as argued in \cite{Tom,Madrid,Madison} --- that achieving a super-Planckian diameter is difficult or impossible.  To correctly assess this, one needs to know how many instantons make relevant contributions to the potential; that is, when is the WGC satisfied by instantons that can be neglected in the potential?

Consider the classical Euclidean action of the $i$th instanton, which takes the form
\be
S^i_\text{CL}={\mathcal S}^i\M \sqrt{({\cal Q}^i) {\bf K}^{-1} ({\cal Q}^i)^\top}\,.
\ee
The prefactor ${\mathcal S}^i$ is fixed for a given class of instantons, e.g.~${\mathcal S}={\sqrt{6}\pi}/{8}$ for gravitational instantons, cf.~(\ref{ginstN}).
To determine the leading instanton(s), it remains to compute $\sqrt{({\cal Q}^i) {\bf K}^{-1} ({\cal Q}^i)^\top}$, which is the invariant length of a charge vector.
We are interested in how small $\sqrt{({\cal Q}^i) {\bf K}^{-1} ({\cal Q}^i)^\top}$ will be in a theory with a given kinetic matrix ${\bf K}$. This will determine the action of the dominant instanton, once ${\mathcal S}$ has been fixed by specifying the class of instantons (Euclidean Dp-brane, gravitational, etc.) under consideration. Naively, one might expect that the typical smallest action is given by
\be\label{minaction1}
\text{Min}\left(S^i_\text{CL}\right)={\mathcal S}{\M\over f_N} \,.
\ee
However, this is in fact just a lower bound on the action. Consider the hypersurface of constant action in a basis where the lattice is ${\mathbb Z}^n$:
\be
 q {\bf K}^{-1} q^\top\le {\rho^2}\,,
 \ee
which defines an ellipsoid ${\cal E}$ with semi-axes of lengths $f_i$. If the eigenvectors of the metric are aligned with the lattice, i.e.~if ${\bf K}$ is diagonal, then the smallest action is in fact given by (\ref{minaction1}). However, for a generic metric the ellipsoid points in some arbitrary direction and may be very elongated. In general, to estimate the smallest action we have to take into account the geometry of that ellipsoid.  Note that the density of the lattice is unity, while the volume of the ellipsoid is given by
 \be
 \text{Vol}({\cal E})={\pi^{N/2}\rho^N\over \Gamma(N/2+1)} \prod_{i=1}^N f_i\,.
 \ee
Thus, we expect that when the volume reaches $\text{Vol}({\cal E})\sim 1$, there will typically exist an integer charge vector $q$ within the ellipsoid. Using Stirling's approximation, we then expect typical actions of the scale
\be
S_{\text{CL}}\sim {\cal S} \M\sqrt{N}\left( \prod_{i=1}^N {1\over f_i}\right)^{1/d}\,.
\ee
That is, we expect the typical scale of the action to be set by the {\it geometric mean} of the eigenvalues, rather than by the largest eigenvalue. While the above is just a heuristic argument to determine the scale of the action, below we will arrive at the same parametric scaling via Minkowski's theorem, which places an upper bound on the minimal action.

\subsection{The dominant instanton and the shortest lattice vector}\label{sec:dominantinstanton}

Computing the minimum size of $\sqrt{({\cal Q}^i) {\bf K}^{-1} ({\cal Q}^i)^\top}$, where the entries of ${\cal Q}^i$ are arbitrary integers, corresponds to solving the shortest vector problem in the charge lattice.
To find a basis of the canonically normalized lattice, we change coordinates to
\be
\boldsymbol \Phi=\diag(f_i)\, \mathbf S^\top_{K} \, \boldsymbol\theta\,,
\ee
where $f_i^2$ are the eigenvalues of the metric ${\mathbf K}$, and the matrix $S^\top_{K}$ diagonalizes the metric
\be
\mathbf S^\top_{K}\, \mathbf K\, \mathbf S^{\phantom{T}}_{K} = \diag(f_i^2)\,.
\ee
The Lagrangian becomes
\be\label{eqn:lagPhi}
{\mathcal L}={1\over 2} \partial \boldsymbol \Phi^\top \partial\boldsymbol\Phi- \sum_{i} \M^4 e^{-S^i}\left[1-\cos\left({\mathscr Q}^i_{\, j}\Phi^j \right) \right]\,,
\ee
where
\be\label{canonicalcharges}
{\mathscr Q}^i={\cal Q}^iS_K \diag(f_i^{-1})\,.
\ee
While all vectors ${\mathscr Q}^i$ define the lattice, we are interested in a {\it reduced} basis of shortest vectors ${\mathscr Q}_{\text{red.}}^i$, where all other elements of ${\mathscr Q}^i$ can be written as integral linear combinations of ${\mathscr Q}_{\text{red.}}^i$ \cite{minkowski1911}. This basis is called {\it{Minkowski reduced}}, and it minimizes the volume of the fundamental parallelepiped:
\be
\det\left({{\mathscr Q}_{\text{red.}}^\top {\mathscr Q}_{\text{red.}}}\right)< \det\left({{\mathscr Q}_{\text{n.red.}}^\top {\mathscr Q}_{\text{n.red.}}}\right)\,,
\ee
where ${\mathscr Q}_{\text{n.red.}}$ is an arbitrary basis of the lattice that is not related to the reduced basis by a unimodular transformation. While in general finding a Minkowski reduced basis is very difficult, Minkowski's theorem sets the scale of the problem by providing an upper bound for the volume of any centrally-symmetric convex set that does not contain a lattice point. In particular, any centrally-symmetric convex set of volume greater than $2^N \det({\mathscr Q}_{\text{red.}})$ contains a lattice point \cite{minkowski1911}.  By considering a sphere of invariant radius $\rho$ we have
\be
{\pi^{N/2}\over \Gamma\left({N\over 2}+1\right)} \rho^N\le 2^N \det({\mathscr Q}_{\text{red.}})\,.
\ee
For the present case, consider the basis ${\mathscr Q}_{\text{red.}}={\cal Q}_{\text{red.}} S_K \diag(f_i^{-1})$, where ${\cal Q}_{\text{red.}}^i$ consists of $N$ vectors of ${\cal Q}^i$ that form a primitive basis. We then have an upper bound for the shortest vector in ${\mathscr Q}_{\text{red.}}$
\be
\text{Min}\left( \lVert{\mathscr Q}_{\text{red.}}^i \rVert\right)\le {\Gamma^{1/N}\left({N\over 2}+1\right)} {2\over{\sqrt{\pi}}}\left(\det\left({\mathscr Q}_{\text{red.}}\right)\right)^{1/N}.
\ee
For the determinant we can write
\be
\det\left({\mathscr Q}_{\text{red.}}\right)=\det\left({\cal Q}_{\text{red.}} S_K \diag(f_i^{-1})\right)=\sqrt{\det({\bf K}^{-1})}\det{{\cal Q}_{\text{red.}}}\,,
\ee
To obtain the most stringent bound, we assume that the matrix ${\cal Q}_{\text{red.}}$ has smallest possible determinant. Because the basis is already primitive, ${\cal Q}_{\text{red.}}$ is a full rank, square, integer matrix, so its determinant is integer, and $|\det({\cal Q}_{\text{red.}})|\ge 1$. Let us assume there exists a matrix ${\cal Q}_{\text{red.}}$ with unit determinant, which gives the most stringent Minkowski bound:
\be\label{minkowskibound}
\text{Min}\left(\left \lVert{\mathscr Q}_{\text{red.}}^i\right \lVert\right)\le {\Gamma^{1/N}\left({N\over 2}+1\right)} {2\over{\sqrt{\pi}}}\left(\prod_{i=1}^N{1\over f_i}\right)^{1/N}\,,
\ee
so that via Stirling's approximation we have
\be\label{minkowskiboundapprox}
\text{Min}\left( \lVert{\mathscr Q}_{\text{red.}}^i \rVert\right)\lesssim  \sqrt{{2N\over e\pi}}\left(\prod_{i=1}^N{1\over f_i}\right)^{1/N}\,.
\ee

While Minkowski's theorem provides an upper bound, a lower bound is easily obtained by assuming that the eigenvectors of ${\bf K}$ are precisely aligned with the faces of the unit hypercube.  In this case we have $\text{Min}_i\left(\left \lVert{\mathscr Q}_{\text{red.}}^i\right \lVert\right)\ge 1/f_N$. However, when considering rotationally invariant  ensembles of metrics ${\bf K}$, the eigenvectors are almost surely not aligned with the faces of the hypercube, and thus the scale of the invariant length of the shortest charge vectors is set by (\ref{minkowskibound}). While the above result is a precise upper bound on the invariant length, we will find below that in the absence of fine-tuning the shortest lattice vector indeed approximately saturates the Minkowski bound.

For the case that all metric eigenvalues are equal, $f_i=f$, we have the constraint
\be
\text{Min}_i\left(\left \lVert{\mathscr Q}_{\text{red.}}^i\right \lVert\right)\le  \sqrt{{2\over e\pi}}{\sqrt{N}\over f}\,,
\ee
which appears to prohibit eigenvalues $f\gtrsim  \sqrt{N}\M$ for instantons with ${\cal S}^i\approx\M$. However, the typical scale of the instanton action is set by the inverse geometric mean of the metric eigenvalues, and so no immediate constraint arises from Minkowski's theorem in the general case where the metric eigenvalues are not identical.
Furthermore, as explained in the introduction, in general the matrix ${\cal Q}_{\text{red.}}$ is not necessarily directly related to the charges that appear in the effective potential that is relevant for inflation, so that in any case there is no direct relation between a constraint on the eigenvalues $f^2_i$ and the diameter of the  fundamental domain relevant to inflation.

\subsection{Instanton actions in large $N$ ensembles}

In the previous section we considered actions of the form
\be
S^i_\text{CL}={\mathcal S}^i \sqrt{({\cal Q}^i) {\bf K}^{-1} ({\cal Q}^i)^\top}\,,
\ee
and argued that, up to the overall scale ${\mathcal S}^i_{\text{CL}}$, the instanton action is bounded by above by Minkowski's theorem. While in general this only is an upper bound, we now examine ensembles of {\it generic} axion theories and demonstrate that the smallest instanton action indeed approximately saturates the bound (\ref{minkowskibound}).

Let us consider a theory with kinetic term
\be
{\mathcal L}_{\text{kin}}={1\over 2}K_{ij} \partial\theta^i\partial\theta^j\,,
\ee
where the lattice consists of the integer vectors ${\mathbb Z}^n$ in $\theta$ coordinates, i.e. ${\cal Q}_{\text{red.}}={\mathbb 1}$, and the metric ${\bf K}$ is chosen from an ensemble that is invariant under orthogonal transformations. In particular, we will consider random matrices ${\bf K}$ that are in the centered Wishart or inverse Wishart ensembles. In order to estimate the leading instanton contribution, up to the overall scale of the action, we are interested in finding the charge vector of least invariant length. In canonically normalized fields, the lattice is defined by the basis
\be
{\mathscr Q}_{\text{red.}}=S_K \diag(f_i^{-1})\,.
\ee
The basis ${\mathscr Q}_{\text{red.}}$ may not be Minkowski reduced.\footnote{Note that the basis ${\mathscr Q}_{\text{red.}}$ is related to the Minkowski reduced basis by a unimodular transformation, so the basis determinant, and hence the resulting bound from Minkowski's theorem, is unchanged.} In general, the problem of finding a Minkowski reduced basis is NP hard and directly maps to a shortest lattice vector problem \cite{Ajtai:1998}.  In this work we employ the LLL algorithm\footnote{We used the Mathematica package by van der Kallen et al.~\cite{lllpackage,cohen1993course}.} to numerically obtain a reduced basis with arbitrary accuracy \cite{Lenstra1982}.

\begin{figure}
  \centering
  \includegraphics[width=1\textwidth]{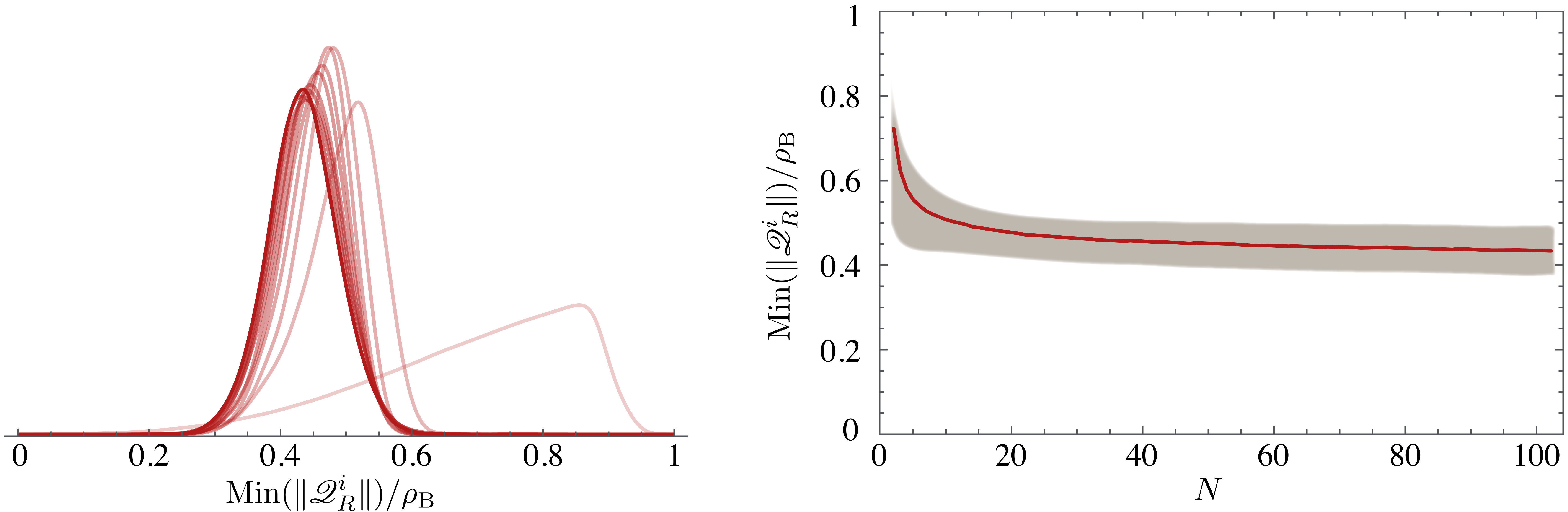}
  \caption{\label{fig:minnormratio}\small Left: Histogram of $\text{Min}_i({\mathscr Q}^i_R)/\rho_{\text{B}}$ for $N=2$ (faintest, rightmost peak) through $N=102$ (boldest, leftmost peak), for the Wishart ensemble. Right: $\langle \text{Min}_i({\mathscr Q}^i_R)/\rho_{\text{B}}\rangle$ over $N$.  Evidently, typical shortest lattice vectors are roughly half the maximum length allowed by the Minkowski bound.}
  \end{figure}

The Minkowski bound is given by (\ref{minkowskibound})
\be
\rho_{\text{B}}={\Gamma^{1/N}\left({N\over 2}+1\right)} {2\over{\sqrt{\pi}}}\left(\prod_{i=1}^N{1\over f_i}\right)^{1/N}\,.
\ee
Figure \ref{fig:minnormratio} shows the distribution of the ratio $\text{Min}_i({\mathscr Q}^i_R)/\rho_{\text{B}}$ for the Wishart ensemble.

For specific metric ensembles where the distribution of eigenvalues is known, we can evaluate the expected Minkowski bound analytically. For example, the determinant of Wishart matrices is distributed as the product of chi-square variables \cite{goodman1963}. Evaluating the expectation value of the resulting product distribution we have
\be
\left\langle\prod_{i=1}^N{f^2_i}\right\rangle=\sigma^{2N}\Gamma(N+1)=\left({f^2_N\over 4 N}\right)^N\Gamma(N+1)\,.
\ee
This gives a typical minimal action of
\be
S^i_{\text{CL}}\sim {\cal S}^i {4\Gamma^{1/N}(1+N/2)\over\sqrt{\pi\Gamma^{1/N}(1+N)}}{\sqrt{N}\M\over f_N} \approx {\cal S}^i\sqrt{8\over \pi}{\sqrt{N}\M\over f_N}\,.
\ee
For the case of inverse Wishart matrices, as motivated for string effective theories in \cite{Long:2014fba}, the typical largest eigenvalue is related to the scale of the entries $\sigma$ by $f_N^2=N/({\calc}  \sigma^2)$, where \cite{BLM,Edelman:1988:ECN:58846.58854,2010arXiv1005.4515M}
\bea
\label{cmedian}
\calc=2+\log(4)-2\sqrt{1+\log(4)}\approx 0.30\,.
\eea

Therefore, we have for metrics in the inverse Wishart ensemble,
\be
S^i_{\text{CL}}\sim {\cal S}^i \sqrt{2\over \calc \pi e^2} {N^{3/2}\over f_N}\,.
\ee
Note that the preceding results give the typical value of the minimal action, i.e.~the value obtained when the metric is a typical member of the corresponding random matrix ensemble.  Considerably larger values of the minimal action are possible if some of the eigenvalues $f_i$ of the metric are atypically small.

\section{Conclusions} \label{conclusions}

In this work we have assessed the limitations imposed on large-field axion inflation by the Weak Gravity Conjecture, and by the contributions of gravitational instantons.

We first clarified that the quantities bounded by the WGC --- including the eigenvalues of a certain charge matrix, cf.~\cite{Madison} --- are nontrivially related to the diameter ${\cal D}$ of the axion fundamental domain, which is the relevant quantity for a putative field range bound excluding large-field inflation (in the absence of monodromy).
In particular, if $f_1^2 \le \ldots \le f_N^2$ are the eigenvalues of the kinetic matrix ${\bf K}$, $f_N <\M$ does not imply that ${\cal D}<\M$, or that ${\cal D}<\sqrt{N}\M$.  The diameter cannot be computed in general from knowledge of the $f_i$ alone: information about the eigenvectors of ${\bf K}$ is necessary in order to determine which terms in the potential define the relevant boundary of the fundamental domain.

We then argued that gravitational instantons do not preclude large-field axion inflation.  We first showed that in a theory of axions coupled to Einstein gravity, gravitational instantons alone suffice to fulfill the zero-form version of the WGC, expressed as the condition that a certain convex hull contains the unit ball.  To determine whether such gravitational instantons necessarily contribute unsuppressed higher harmonics to the potential, we used Minkowski's theorem to estimate the smallest gravitational instanton action, which is specified by the shortest vector in the charge lattice.  In this problem as well, information about the orientation of the eigenvectors of ${\bf K}$ was crucial.  We found that in generic large $N$ theories, $S_{\rm{min}} \gtrsim \sqrt{N}\M/f_N$, and so gravitational instanton contributions are under good control even for $f_N =\M$.

We conclude that in theories of many axions, the limitations obtained to date from the WGC, and from gravitational instantons, do not exclude axion displacements as large as the upper bound set by observational limits on the tensor-to-scalar ratio \cite{Ade:2015tva}.

Even so, it would be valuable to give a more precise characterization of quantum gravity constraints on axion inflation.  As we have shown in this work, moving from single-axion theories to generic many-axion theories reveals nontrivial alignment phenomena that deserve continued exploration.  Moreover, although the example of a flux compactification with a Planckian diameter \cite{DDFGK} that we analyzed in \cite{BLM} obeys the WGC condition on the quantities denoted $f_i$ in \cite{Madison}, we have not determined whether the complete set of instantons in that theory fulfills the convex hull condition required by the WGC.  More generally, understanding how super-Planckian diameters in string theory might be limited by the WGC remains an important question.
\section*{Acknowledgements}
We thank Tom Hartman and John Stout for useful discussions and Tom Rudelius  for correspondence on related topics. This work was supported by NSF grant PHY-0757868.

\bibliographystyle{modifiedJHEP}
\bibliography{refsWGC}

\providecommand{\href}[2]{#2}\begingroup\raggedright\begin{thebibliography}{10}

\bibitem{Tom}
T.~Rudelius, ``{Constraints on Axion Inflation from the Weak Gravity
  Conjecture},'' \href{http://xxx.lanl.gov/abs/1503.0079}{{\tt
  arXiv:1503.0079}}.

\bibitem{Madrid}
M.~Montero, A.~M. Uranga, and I.~Valenzuela, ``{Transplanckian axions !?},''
  \href{http://xxx.lanl.gov/abs/1503.0388}{{\tt arXiv:1503.0388}}.

\bibitem{Madison}
J.~Brown, W.~Cottrell, G.~Shiu, and P.~Soler, ``{Fencing in the Swampland:
  Quantum Gravity Constraints on Large Field Inflation},''
  \href{http://xxx.lanl.gov/abs/1503.0478}{{\tt arXiv:1503.0478}}.

\bibitem{Natural}
K.~Freese, J.~A. Frieman, and A.~V. Olinto, ``{Natural inflation with pseudo -
  Nambu-Goldstone bosons},'' Phys.Rev.Lett. {\bf 65} (1990) 3233--3236.

\bibitem{SW}
E.~Silverstein and A.~Westphal, ``{Monodromy in the CMB: Gravity Waves and
  String Inflation},'' Phys.Rev. {\bf D78} (2008) 106003,
  [\href{http://xxx.lanl.gov/abs/0803.3085}{{\tt arXiv:0803.3085}}].

\bibitem{MSW}
L.~McAllister, E.~Silverstein, and A.~Westphal, ``{Gravity Waves and Linear
  Inflation from Axion Monodromy},'' Phys.Rev. {\bf D82} (2010) 046003,
  [\href{http://xxx.lanl.gov/abs/0808.0706}{{\tt arXiv:0808.0706}}].

\bibitem{Banks:2003sx}
T.~Banks, M.~Dine, P.~J. Fox, and E.~Gorbatov, ``{On the possibility of large
  axion decay constants},'' JCAP {\bf 0306} (2003) 001,
  [\href{http://xxx.lanl.gov/abs/hep-th/0303252}{{\tt hep-th/0303252}}].

\bibitem{ArkaniHamed:2006dz}
N.~Arkani-Hamed, L.~Motl, A.~Nicolis, and C.~Vafa, ``{The String landscape,
  black holes and gravity as the weakest force},'' JHEP {\bf 0706} (2007) 060,
  [\href{http://xxx.lanl.gov/abs/hep-th/0601001}{{\tt hep-th/0601001}}].

\bibitem{BLM}
T.~C. Bachlechner, C.~Long, and L.~McAllister, ``{Planckian Axions in String
  Theory},'' \href{http://xxx.lanl.gov/abs/1412.1093}{{\tt arXiv:1412.1093}}.

\bibitem{DDFGK}
F.~Denef, M.~R. Douglas, B.~Florea, A.~Grassi, and S.~Kachru, ``{Fixing all
  moduli in a simple f-theory compactification},'' Adv.Theor.Math.Phys. {\bf 9}
  (2005) 861--929, [\href{http://xxx.lanl.gov/abs/hep-th/0503124}{{\tt
  hep-th/0503124}}].

\bibitem{Cheung:2014vva}
C.~Cheung and G.~N. Remmen, ``{Naturalness and the Weak Gravity Conjecture},''
  Phys.Rev.Lett. {\bf 113} (2014) 051601,
  [\href{http://xxx.lanl.gov/abs/1402.2287}{{\tt arXiv:1402.2287}}].

\bibitem{ArkaniHamed:2007js}
N.~Arkani-Hamed, J.~Orgera, and J.~Polchinski, ``{Euclidean wormholes in string
  theory},'' JHEP {\bf 0712} (2007) 018,
  [\href{http://xxx.lanl.gov/abs/0705.2768}{{\tt arXiv:0705.2768}}].

\bibitem{KNP}
J.~E. Kim, H.~P. Nilles, and M.~Peloso, ``{Completing natural inflation},''
  JCAP {\bf 0501} (2005) 005,
  [\href{http://xxx.lanl.gov/abs/hep-ph/0409138}{{\tt hep-ph/0409138}}].

\bibitem{Nflation}
S.~Dimopoulos, S.~Kachru, J.~McGreevy, and J.~G. Wacker, ``{N-flation},'' JCAP
  {\bf 0808} (2008) 003, [\href{http://xxx.lanl.gov/abs/hep-th/0507205}{{\tt
  hep-th/0507205}}].

\bibitem{Bachlechner:2014hsa}
T.~C. Bachlechner, M.~Dias, J.~Frazer, and L.~McAllister, ``{A New Angle on
  Chaotic Inflation},'' \href{http://xxx.lanl.gov/abs/1404.7496}{{\tt
  arXiv:1404.7496}}.

\bibitem{minkowski1911}
H.~Minkowski, {\em Gesammelte Abhandlungen von Hermann Minkowski: Vol.: 2}.
\newblock B.G. Teubner, 1911.

\bibitem{Ajtai:1998}
M.~Ajtai, {\it The shortest vector problem in l2 is np-hard for randomized
  reductions (extended abstract)},  in {\em Proceedings of the Thirtieth Annual
  ACM Symposium on Theory of Computing}, STOC '98, (New York, NY, USA),
  pp.~10--19, ACM, 1998.

\bibitem{lllpackage}
W.~van~der Kallen, ``Implementations of extended lll,''
  \url{http://www.staff.science.uu.nl/~kalle101/lllimplementations.html}, 1998.

\bibitem{cohen1993course}
H.~Cohen, {\em A Course in Computational Algebraic Number Theory}.
\newblock Graduate Texts in Mathematics. Springer, 1993.

\bibitem{Lenstra1982}
A.~Lenstra, H.~Lenstra, and L.~Lov\'{a}sz, ``Factoring polynomials with
  rational coefficients,'' Mathematische Annalen {\bf 261} (1982), no.~4
  515--534.

\bibitem{goodman1963}
N.~R. Goodman, ``The distribution of the determinant of a complex wishart
  distributed matrix,'' Ann. Math. Statist. {\bf 34} (03, 1963) 178--180.

\bibitem{Long:2014fba}
C.~Long, L.~McAllister, and P.~McGuirk, ``{Heavy Tails in Calabi-Yau Moduli
  Spaces},'' JHEP {\bf 1410} (2014) 187,
  [\href{http://xxx.lanl.gov/abs/1407.0709}{{\tt arXiv:1407.0709}}].

\bibitem{Edelman:1988:ECN:58846.58854}
A.~Edelman, ``Eigenvalues and condition numbers of random matrices,'' SIAM J.
  Matrix Anal. Appl. {\bf 9} (Dec., 1988) 543--560.

\bibitem{2010arXiv1005.4515M}
S.~N. {Majumdar}, ``{Extreme Eigenvalues of Wishart Matrices: Application to
  Entangled Bipartite System},'' ArXiv e-prints (May, 2010)
  [\href{http://xxx.lanl.gov/abs/1005.4515}{{\tt arXiv:1005.4515}}].

\bibitem{Ade:2015tva}
{\bf BICEP2, Planck} Collaboration, P.~Ade {\em et.~al.}, ``{Joint Analysis of
  BICEP2/$Keck Array$ and $Planck$ Data},'' Phys.Rev.Lett. {\bf 114} (2015),
  no.~10 101301, [\href{http://xxx.lanl.gov/abs/1502.0061}{{\tt
  arXiv:1502.0061}}].

\end{thebibliography}\endgroup
\end{document}